\documentclass[aps,pra,letterpaper,nobalancelastpage,twocolumn,superscriptaddress,floatfix,nofootinbib]{revtex4-2}

\usepackage{lineno}
\usepackage{graphicx}
\usepackage{amsmath}
\usepackage{bbold}
\usepackage{amssymb}
\usepackage[english]{babel}
\usepackage{color}
\usepackage[version=4]{mhchem}
\usepackage[hidelinks]{hyperref}
\hypersetup{colorlinks=true,citecolor=blue,linkcolor=blue,urlcolor=blue,pdfstartview=FitH}
\usepackage{dsfont}
\usepackage{placeins}
\usepackage{mathtools}
\usepackage[normalem]{ulem}
\usepackage{lipsum}
\usepackage{comment}
\usepackage{booktabs}
\usepackage{mathtools}
\usepackage{array}
\usepackage{tabularx}

\usepackage{multirow}

\renewcommand{\selectlanguage}[1]{}

\setcitestyle{super}

\usepackage{soul}
\newcommand{\Caltech}{California Institute of Technology, Pasadena, CA 91125, USA}
\newcommand{\Oratomic}{Oratomic, Pasadena, CA 91125, USA}

\usepackage{caption}
\usepackage[caption=false]{subfig}

\DeclareCaptionLabelSeparator{bar}{ \textbf{\textbar}~}
\usepackage{booktabs,tabularx,array}
\newcolumntype{Y}{>{\centering\arraybackslash}X}
\usepackage{enumitem}
\setlist{nolistsep}
\begin{document}
\nocite{myControl}

\title{Local polarization control of optical tweezer arrays}

\author{Xiangkai Sun}
\thanks{These authors contributed equally to this work.}
\author{Richard Bing-Shiun Tsai}
\thanks{These authors contributed equally to this work.}
\author{Andrew Winnicki}
\author{Yuan Le}
\author{Kon H. Leung}
\author{Nelson Darkwah Oppong}
\affiliation{\Caltech}
\author{Manuel Endres}
\email{mendres@caltech.edu}
\affiliation{\Caltech}
\affiliation{\Oratomic}

\begin{abstract}
    Optical tweezer arrays trapping ultracold atoms and molecules are a versatile quantum science platform with broad impact in quantum simulation, quantum computation, and quantum metrology. The polarization dependence of anisotropic vector and tensor light shifts offers a degree of freedom for precise quantum state engineering of trapped particles. Similarly, state-selective operations depend on the polarization of the addressing light through the atom--light coupling strength. Yet, methods for controlling the polarization of individual tweezers or local addressing beams remain an experimental challenge. Here, we demonstrate independent, site-resolved linear polarization rotation across an optical tweezer array by exploiting the local birefringence tunability of a spatial light modulator. Applying this capability to an array of $^{88}\rm{Sr}$ atoms, we homogenize differential light shifts imparted by the 813-nm tweezers to the narrow ${^1{\rm S}_0}\leftrightarrow{^3{\rm P}_1}$ transition, enabling sideband cooling under a magic-angle condition. Furthermore, we show dynamic transport of trapped atoms across distinct polarization zones with high survival and preserved coherence. Finally, we characterize tweezer polarization noise and demonstrate closed-loop stabilization referenced to the atomic transition frequency, reaching mrad-level stability. Our work establishes a technique for manipulating polarization-sensitive atomic and molecular transitions in tweezer architectures, with immediate applications to optical tweezer clocks and multi-zone quantum processors.
\end{abstract}

\maketitle

\section{Introduction}

Optical tweezer arrays are a powerful platform for assembling and controlling ultracold matter at the single-particle level, enabling advances in quantum simulation~\cite{browaeys_many-body_2020,bernien_probing_2017,semeghini_probing_2021,scholl_quantum_2021,young2022tweezer,evered2025probing,sun2026observation}, fault-tolerant quantum computation~\cite{Bluvstein2026,atomcomputing2026}, and quantum metrology~\cite{Finkelstein2024,Cao2024,bornet2023scalable}. Typically, these arrays are generated by spatially multiplexing a single global laser beam~\cite{Barredo2016,Endres2016,wang_preparation_2020,Schymik2022,Manetsch2025,holman_trapping_2026}, with most approaches affording considerable in-situ control over the position and intensity of each tweezer. In contrast, the tweezer polarization is often globally controlled, and optical elements in the beam path may introduce parasitic polarization errors.

For states with anisotropic polarizabilities, the intense trapping light causes polarization-dependent transition frequency shifts~\cite{LeKien2013,thompson_coherence_2013}. These shifts impact fine-structure qubits in divalent atoms~\cite{Unnikrishnan2024,Tao2026,Ammenwerth2025}, narrow-line cooling of lanthanides~\cite{bloch2023trapping,grun2024optical}, cooperative emission in subwavelength atom arrays~\cite{holman2026mid,holman_trapping_2026}, and rotational coherence in polar molecules~\cite{Burchesky2021,Park2023,gregory2024second}. In parallel, laser polarization sets the atom--light coupling strength for driving local single-qubit gates on hyperfine~\cite{levine2022dispersive} or nuclear-spin~\cite{muniz2025high,ma2022universal,lis2023midcircuit,nakamura2024hybrid} qubits in quantum processors~\cite{Bluvstein2024,Bluvstein2026,zhang2026logical}. For spatially small tweezer arrays, tuning the polarization of the global trap laser has proven sufficient to realize magic trapping conditions (i.e. the same trapping potential for ground and excited state), both for cooling~\cite{Cooper2018,Norcia2018,bloch2023trapping,grun2024optical} and for long internal-state coherence~\cite{Burchesky2021,Park2023,Picard2025,Ammenwerth2025}. However, this approach has limited scalability. In arrays with large spatial extent, the wide angular span of the beam produces site-to-site polarization distortions that cannot be undone by global adjustments, necessitating local polarization control.

Here, we demonstrate independently programmable polarization rotation at the individual sites of an optical tweezer array. To this end, we use a liquid-crystal spatial light modulator (SLM) at an intermediate image plane in the pathway of 813-nm tweezers that trap a one-dimensional array of ${}^{88}\mathrm{Sr}$ atoms. The SLM functions as a birefringent phase retarder with high spatial resolution, enabling us to control the polarization of each tweezer individually. We program polarization patterns onto the tweezer array in two configurations: {(i)} a uniform polarization pattern that sets every site to the tensor magic-angle condition for the ${}^1{\rm S}_0\leftrightarrow{}^3{\rm P}_1$ intercombination transition, and {(ii)} a staggered pattern in which two groups of atoms are spectrally isolated. We explore elements of a polarization-defined zone architecture, comprising storage, cooling, and computation zones, which can be addressed selectively by laser frequency.. As a step in this direction, we demonstrate sideband cooling by engineering magic-angle conditions with local polarization tuning. Additionally, we show coherent transport between distinct polarization zones. Finally, we characterize polarization noise in detail using both optical polarimetry and atomic spectroscopy, finding mrad-level polarization noise and drifts that closed-loop atomic feedback further suppresses on long timescales. Our work establishes local polarization control for tweezer arrays or addressing setups as a new tool for quantum science applications.

\section{Experimental Setup}\label{sec:setup}
\begin{figure}[t]
    \centering
    \includegraphics[width=1.0\linewidth]{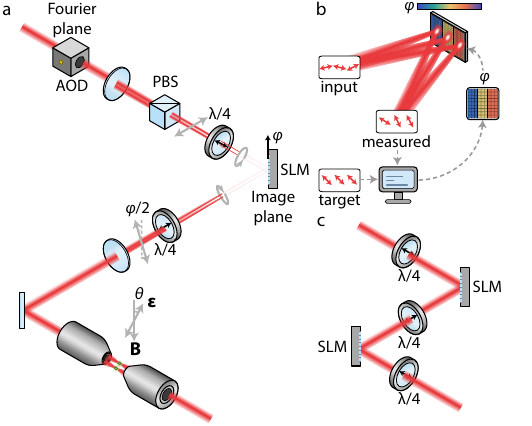}
    \caption{\textbf{Programming of individual tweezer polarization.}
    \textbf{(a)} Setup for local polarization control. An acousto-optic deflector (AOD) generates a one-dimensional array that is relayed to an image plane inside the vacuum chamber to create optical tweezer traps. An SLM placed at an intermediate image plane, combined with a pair of QWPs (oriented at $45^\circ$ to the SLM's principal axes), allows for arbitrary linear polarization rotations of each tweezer. The rotation is half the phase $\varphi$ imprinted by the illuminated SLM pixels.
    \textbf{(b)} SLM phase pattern at the intermediate image plane. The SLM mask is divided into ``superpixels'' (shown as differently colored regions) with a single phase applied per superpixel. Measured tweezer polarizations (from atomic or optical signals) are compared with the targets, and the resulting phase correction is applied to the mask.
    \textbf{(c)} Proposed setup for full polarization control (relay optics omitted). Two SLMs and three QWPs afford independent control of both ellipticity (upstream SLM) and principal-axis orientation (downstream SLM).}
    \label{fig:optics}
\end{figure}

The experimental setup is depicted in Fig.~\ref{fig:optics}a. A global 813-nm laser beam is diffracted by an acousto-optic deflector (AOD) driven by a multi-tone radio frequency input and linearly polarized by a polarizing beam splitter. The multi-tone drive of the AOD generates a one-dimensional tweezer array~\cite{Endres2016}. The AOD diffraction is optically relayed to a microscope objective that focuses the beams into an array of optical tweezers at the atom (i.e., image) plane inside the vacuum chamber. To realize independent linear polarization rotations for the tweezers, we place an SLM (Meadowlark Optics E19x12-500-1200-HDM8, $1200 \times 1920 \,\rm{px}^2$ resolution) at an intermediate image plane downstream of the AOD. A quarter waveplate (QWP) placed before the SLM converts the initial linear polarization into an elliptical state. The SLM tunes this ellipticity by imparting a phase $\varphi$ on the focused spot, which a subsequent QWP converts into a linear polarization rotation of $\varphi/2$. Consequently, we achieve control over the angle $\theta$ between each tweezer's polarization axis and the applied magnetic field ($\approx450~{\rm G}$, aligned within the atom plane).

In the experiment, the atoms are trapped in a 13-site one-dimensional (1D) array (trap depth $\approx k_B\times 0.5~{\rm mK}$). Details related to the strontium tweezer array are reported in our previous works~\cite{Cooper2018,Covey2019A,Madjarovthesis}. The tweezers are spaced in the atom plane by $6.6~{\rm \mu m}$ and have a $1/e^2$ waist radius of $0.8~{\rm \mu m}$. The demagnification between the SLM and atom planes is approximately $100$. With an SLM pixel size of $8.0~{\rm \mu m}$, these parameters correspond, at the SLM plane, to a tweezer spacing of $82$ pixels and a waist of $w\approx 10$ pixels. To write a spatially uniform phase across the tweezer beam profile, we partition the SLM mask into rectangular regions ($82\times 1920$ pixels) to form superpixels; i.e., the superpixel width along the array axis is $d\approx82$ pixels and $d/w\approx 8$ (Fig.~\ref{fig:optics}b). A phase value for each superpixel is then computed (Sec.~\ref{sec:local}) and sent to the SLM controller to realize the target polarizations. 
For most measurements in this work, the width of the superpixels equals the tweezer spacing on the intermediate imaging plane. In additional tests (not shown), we kept the tweezer spacing fixed and reduced the width of the independently programmable superpixels centered on each tweezer, while holding the intervening pixel regions at a fixed phase. We find that a superpixel width of $30$ pixels ($d=3w$), containing $>99\%$ of the tweezer beam power, is sufficient for independent polarization control, with no appreciable degradation due to potential electric-field crosstalk between neighboring SLM pixels. Smaller superpixel widths may also be usable, although we did not establish a strict minimum. We discuss the implications for scalability to larger arrays in the outlook.

Although it lies beyond the scope of the present work, we note that a straightforward extension of our method permits arbitrary synthesis of pure polarization states. Incorporating a second SLM and a third QWP into the setup described above (Fig.~\ref{fig:optics}c and Appendix~\ref{Appendix:Jones}) allows both ellipticity and principal-axis orientation to be tuned independently at each site.

\section{Local polarization control}\label{sec:local}
\begin{figure}[t]
    \centering
    \includegraphics[width=1.0\linewidth]{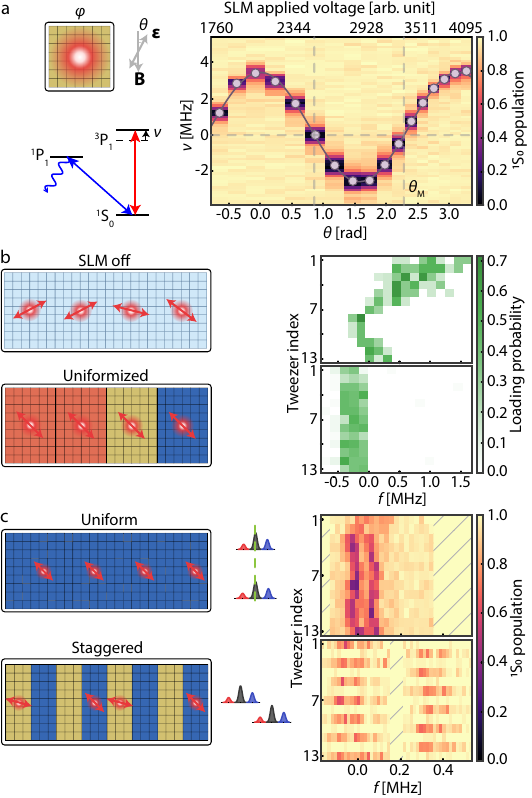}
    \caption{\textbf{Local control of atomic transition frequencies.}
    \textbf{(a)} Polarization tuning of the tensor light shift. We apply a voltage to a superpixel and measure the differential light shift, $\nu$, of the ${}^1{\rm S}_0\leftrightarrow{}^3{\rm P}_1$ transition relative to free space, allowing us to extract $\theta$, the angle between the linear tweezer polarization $\epsilon$ and the bias magnetic field $B$.
    \textbf{(b)} Tweezer loading is signaled by successful atom imaging, which in turn requires that the same global cooling-laser detuning be effective at every site. Left uncompensated, site-dependent light shifts cause a large scatter in the optimal cooling laser frequencies (top). With local polarization correction, a single global cooling frequency loads the entire array efficiently (bottom).
    \textbf{(c)} We program the SLM superpixels to produce either a uniform (top) or alternating (bottom) pattern of line centers across the array. Two distinct resonances are apparent in every scan: the carrier transition and the first blue sideband. Regions that do not contain data are masked for visual clarity (diagonally shaded). In (b) and (c), $f$ denotes the stepped laser frequency offset relative to the free-space resonance.}
    \label{fig:polarization}
\end{figure}

Having established the SLM-based polarization-rotation scheme, we now calibrate the relationship between experimentally inferred $\theta$ and the applied voltage on the liquid crystal pixels. To this end, we perform spectroscopy on the narrow intercombination transition (${}^1{\rm S}_0\leftrightarrow{}^3{\rm P}_1$, natural linewidth $\Gamma \approx 2\pi \times 7~{\rm kHz}$) with a single trapped atom and observe the differential light shifts, $\nu$, as a function of the applied pixel voltages. As shown in Fig.~\ref{fig:polarization}a, the solid curve is a fit of the function $\nu = \Delta_{\rm LS}^s - \Delta_{\rm LS}^t (3\cos^2\theta - 1)$ to the data, with $\theta$ expressed as a polynomial function of the applied voltage (Appendix~\ref{Appendix:SLMcalib}). The fit assumes that the atom experiences linearly polarized light. The coefficients of the polynomial, along with $\Delta_{\rm LS}^{(s/t)}$ (which encode the polarization-independent scalar light shift and the polarization-dependent tensor light shift, respectively), are fit parameters. As expected, the SLM enables a full $\theta \in [0,\pi]$ rotation of the tweezer polarization axis (Fig.~\ref{fig:polarization}a).

Atom temperature crucially determines the tweezer loading probability, imaging survival, and optical-qubit coherence~\cite{Finkelstein2024,Tsai2026}. Sisyphus cooling is a reliable technique for preparing ultracold samples of strontium atoms confined in deep non-magic potentials~\cite{Covey2019A,Tao2024}. However, the efficacy of Sisyphus cooling is contingent on suppressing the array intensity inhomogeneity to the level of ${<}1\%$ (i.e., reducing the tweezer-induced light shift inhomogeneity to be comparable with $\Gamma$). 
For wide-angle two-dimensional (2D) crossed-AOD-generated arrays, we find it challenging to reduce the intensity inhomogeneity below $2\%$~\cite{Cooper2018}. These inhomogeneities originate in polarization distortions resulting from highly non-paraxial incidence on optical elements---downstream polarization discriminators (e.g., polarizing beam splitters) then convert these polarization errors to intensity variations. Furthermore, a pair of crossed AODs uses $2N$ independently controllable tones to create an $N \times N$ array, which provides too few degrees of freedom for control of individual tweezers.

The difficulty is illustrated by the tweezer loading probability as a function of cooling laser frequency (Fig.~\ref{fig:polarization}b, top). Left uncompensated and without a clean-up polarizing element, these polarization inhomogeneities cause the frequency for optimal loading to vary by ${>}1~{\rm MHz}$ across sites. To achieve efficient cooling in this scenario, we instead individually tune the tweezer polarization to the magic-angle condition~\cite{Norcia2018,Cooper2018}. Under this condition, the state-dependent differential light shifts vanish ($\nu=0$), decoupling the sensitivity of the cooling resonance from site-to-site tweezer intensity non-uniformity. We locate the magic angle ($\theta_M$) experimentally (Fig.~\ref{fig:polarization}a), finding that differential trap light shifts are nulled at $\theta = \theta_M = 49.2(4)^\circ$. By applying in-situ corrections to $\theta$ for each tweezer, we find that the optimal loading condition collapses to a single cooling-laser frequency across all traps (Fig.~\ref{fig:polarization}b, bottom).

To investigate the frequency uniformity of the corrected array in greater detail, we probe the spread of the ${}^1{\rm S}_0\leftrightarrow{}^3{\rm P}_1$ line centers, for which the angles $\theta$ are (ideally) tuned to the magic angle at every site. After one to three rounds of iterative corrections, we typically achieve a transition frequency standard deviation of 10--20~${\rm kHz}$ (Fig.~\ref{fig:polarization}c, top). For our trap depth ($\approx$0.5~mK), the derivative $\partial \nu/\partial\theta$ at the magic-angle condition is $5.95(8)~\rm{kHz/mrad}$. Hence, we infer a linear polarization inhomogeneity of 2--3~${\rm mrad}$, comparable to the $3~{\rm mrad}$ discrete rotation step (integer increment by two on the applied voltage $V$, Appendix~\ref{Appendix:SLMcalib}) set by the phase resolution of the SLM controller. The stability of the programmed polarizations is further discussed in Sec.~\ref{sec:stability}.

Our technique offers more than uniformization, allowing deterministic programmability of tweezer polarization rotations in different spatial regions of the atom plane. To illustrate this, we create a polarization pattern that facilitates magic-angle sideband cooling on odd-index tweezers while shifting the resonance of even-index tweezers $0.4~{\rm MHz}$ away (Fig.~\ref{fig:polarization}c, bottom). Notably, the spectral resonances are fully non-overlapping, indicating the possibility of selectively addressing a subset of atoms, an essential ingredient of the polarization-augmented zone architecture, which we detail in the following section.

\section{Zone architecture}\label{sec:zone}

\begin{figure*}[t]
    \centering
    \includegraphics[width=1.0\linewidth]{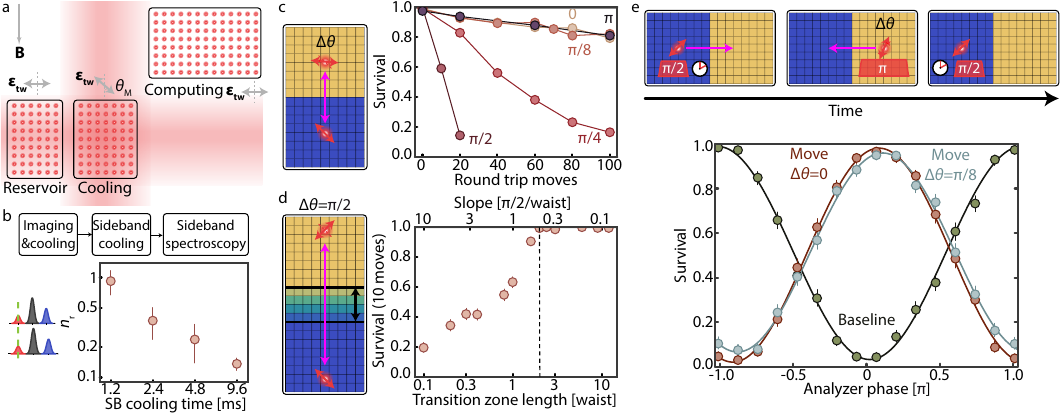}
    \caption{\textbf{Zone architecture for optical tweezer arrays with local polarization programmability.}
    \textbf{(a)} Proposed zone layout for a 2D array. Three orthogonal cooling beams intersect to form the cooling zone, where the tweezer polarizations are tuned to $\theta \approx \theta_M$. Atoms in the reservoir and computing zones are spectrally isolated from the cooling light via tweezer polarization tuning.
    \textbf{(b)} Magic-angle sideband cooling on the ${}^1{\rm S}_0\leftrightarrow{}^3{\rm P}_1$ transition following fluorescence imaging, achieving a site-averaged radial motional occupation of $\bar{n}_r = 0.137(19)$ after 9.6~ms of cooling time.
    \textbf{(c)} Inter-zone transport survival versus total polarization change $\Delta\theta$. We observe the onset of atom loss for $|\Delta\theta| > \pi/8$, which we attribute to tweezer mode deformation at the sharp boundary between zones. The survival fully recovers at $|\Delta\theta| = \pi$, where the phase $\Delta \varphi = 2\pi$ wraps around.
    \textbf{(d)} A linear phase ramp bridges the $|\Delta\theta| = \pi/2$ polarization change between two superpixels. An identical move for a uniform phase mask serves as the baseline survival reference. Survival improves with gentler transition zones (longer ramp lengths), recovering the baseline for ramp lengths $\gtrsim 2w$ (dashed vertical line).
    \textbf{(e)} Coherence-preserving transport across polarization-defined zones. The spin-echo coherence after an 800-$\mu$s round trip is 87(2)\% for $|\Delta\theta| = \pi/8$, versus 96(2)\% under a uniform phase mask.}
    \label{fig:zone}
\end{figure*}

In Fig.~\ref{fig:zone}a, we envision a zone-based 2D architecture for a strontium quantum processor that leverages local polarization programmability. In the cooling zone, atoms are trapped in AOD-generated tweezers tuned to $\theta \approx \theta_M$ and sideband cooled to the motional ground state. Subsequently, the atoms are transported by the same tweezers to a computing zone for quantum information processing. Outside of the cooling zone, polarization rotation spectrally detunes the ${}^1{\rm S}_0\leftrightarrow{}^3{\rm P}_1$ transition away from resonance with the cooling laser, isolating those atoms from the cooling light and minimizing crosstalk. Below, we demonstrate the key building blocks of this architecture. 

\subsection{Magic-Angle Sideband Cooling}

We first detail sideband cooling on the ${}^1{\rm S}_0\leftrightarrow{}^3{\rm P}_1$ transition at the magic-angle condition, utilizing our technique of local polarization tuning (see Sec.~\ref{sec:local}). Experimentally, we employ three independent beams aligned to the tweezers: two drive radial motional transitions, while the third is delivered through the microscope objective to couple to the axial mode. The cooling beams possess the appropriate polarization component to drive $\pi$-transitions---at our applied magnetic field strength, the other Zeeman sublevels are shifted out of resonance by $>$900 MHz. At a saturation parameter of ${\approx}60$ and a Lamb-Dicke parameter of $\approx 0.2$, the carrier transition and the red-sideband transition to the ground state are power broadened to approximately $8\Gamma$ and $2\Gamma$, respectively. The optimization of the cooling parameters is discussed in Appendix~\ref{sec:sideband}.

To illustrate the cooling performance under realistic experimental conditions, where atomic heating occurs predominantly during fluorescence imaging~\cite{Covey2019A}, we apply cooling simultaneously with imaging, followed by additional cooling cycles. After these cooling cycles, we precisely quantify the radial motional occupation, $n_r$, by probing the sideband asymmetry on the ultra-narrow clock transition (${}^1{\rm S}_0\leftrightarrow{}^3{\rm P}_0$). As shown in Fig.~\ref{fig:zone}b, with 40 cycles ($9.6~{\rm ms}$) of sideband cooling, we recool the atoms to a site-averaged $\bar n_r = 0.137(19)$. This cooling is faster and reaches a lower final occupation than Sisyphus cooling of Sr atoms in optical tweezers with trapping frequencies comparable to ours~\cite{Covey2019A}. Although this is a proof-of-concept demonstration on a 13-site 1D array, the results suggest a straightforward extension to 2D architectures.

\subsection{Atom Transport Across Polarization Zones}
We now turn to atom transport between zones of differing tweezer polarization. We explore techniques that preserve atom survival as well as the coherence of clock qubits.

We first investigate how a sudden change in tweezer polarization impacts atom transport survival. To this end, we program a bipartite profile featuring a sharp polarization change of $|\Delta \theta|$ between adjacent zones (Fig.~\ref{fig:zone}c). We perform 100 round trips---crossing the boundary twice per trip, with a one-way distance of $3.3~{\rm \mu m} \approx 4w$---and image the atoms at the end of the sequence without recooling. The flat profile ($|\Delta \theta| = 0$) sets the baseline loss attributable to transport alone. We observe no measurable degradation in survival for polarization changes up to $|\Delta \theta| = \pi/8$. However, when $|\Delta \theta|$ increases from $\pi/8$ to $\pi/2$, atom survival drops markedly. We attribute this behavior to spatial mode deformation of the tweezer at the sharp interface, corroborated by direct beam-profile images of the tweezers. This mode deformation may be readily understood by considering the instant at which the tweezer straddles the phase boundary of two superpixels with $\Delta \varphi = \pi$ (corresponding to $|\Delta \theta| = \pi/2$). Here, the two halves of the tweezer acquire orthogonal polarizations and no longer interfere to reproduce the original trap at the atom plane; instead, each half undergoes knife-edge diffraction. Atom loss thus stems from motional heating driven by the time-dependent modulation of the potential as the tweezer crosses the boundary. Notably, at $|\Delta \theta|=\pi$ the phase wraps fully ($\Delta \varphi = 2\pi$), and the survival accordingly returns to baseline. 

To overcome this difficulty, we introduce a linear phase ramp between two superpixels to smooth the polarization transition over a finite distance. Restricting our attention to the most adverse case ($|\Delta \theta| = \pi/2$), we characterize atom survival after 10 round trips with a one-way move distance of $13.3~{\rm \mu m}$ ($17w$) for various transition zone lengths (Fig.~\ref{fig:zone}d). As expected, the survival improves for longer ramp lengths (i.e., gentler slopes). Favorably, we find that we recover the $\Delta \theta = 0$ baseline for ramp lengths $\gtrsim 2.0w$, corresponding to changing the polarization by $0.25 \pi$ per tweezer waist. Since $|\Delta\theta| = \pi/2$ is the most adverse case, the result suggests that a transition distance of $2.0w$ between any two polarization zones suffices as an operational minimum. 

Finally, we investigate clock-qubit (${}^1{\rm S}_0 \leftrightarrow {}^3{\rm P}_0$) coherence when atoms are shuttled across distinct polarization domains. We implement a spin-echo sequence of total duration $0.8~{\rm ms}$ and compare the fringe contrast for three situations (Fig.~\ref{fig:zone}e): (i) a static atom serving as the baseline (dark green), (ii) an atom moved under a uniform phase mask ($|\Delta \theta| = 0$, red), and (iii) an atom moved across a sharp bipartite step ($|\Delta \theta| = \pi/8$, light green). Transport under a uniform phase mask leaves the contrast at $96(2)\%$, comparable to the static-atom baseline. However, moving the atom across the abrupt polarization change lowers the contrast to $87(2)\%$. Since the clock transition is insensitive to tweezer polarization (both states possess zero total electronic angular momentum), we attribute this contrast reduction to atom heating during transport. As shown earlier, smoothing the polarization transition restores the baseline survival even for $|\Delta\theta| = \pi/2$, and we expect this approach to suppress heating as well. We leave an explicit experimental test to future work.

\section{Polarization noise and stabilization}\label{sec:stability}
\begin{figure}[t]
    \centering
    \includegraphics[width=1.0\linewidth]{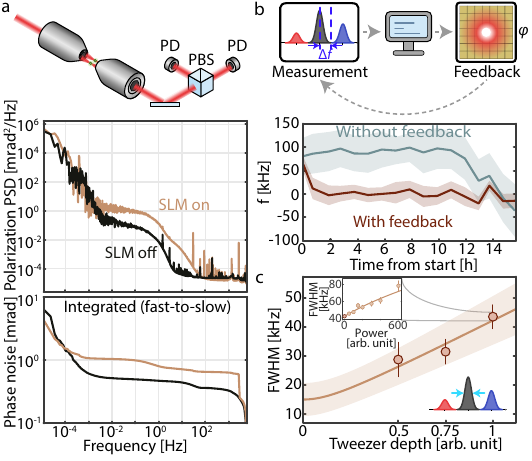}
    \caption{\textbf{Polarization noise and stabilization.}
    \textbf{(a)} Top: polarimeter comprising a polarizing beam splitter (PBS) and two photodiodes (PDs). Bottom: power spectral density (PSD) of the linear polarization-angle noise. When used for local tweezer polarization rotation, the SLM introduces significant excess polarization-angle noise in the $10^{-3}$~Hz to $10^{4}$~Hz band. Over a 24-hour measurement, the integrated polarization noise is $\approx$5~mrad, dominated by slow drifts correlated with ambient humidity and temperature.
    \textbf{(b)} On hour-long timescales, we actively stabilize the tweezer polarizations by updating the superpixel phases to steer the probed line centers of the ${}^1{\rm S}_0 \leftrightarrow {}^3{\rm P}_1$ carrier transition to the free-space value. Solid red (black) curves show the average line centers of sites in closed (open) loop; shaded bands represent the standard deviation over sites. The standard deviation over 16 hours is 20~kHz with feedback and 44~kHz without, corresponding to 3.3~mrad and 7.3~mrad in polarization angle, respectively.
    \textbf{(c)} ${}^1{\rm S}_0 \leftrightarrow {}^3{\rm P}_1$ linewidths versus trap depth near the magic angle, each extrapolated to zero probe-laser power (inset). Polarization-angle noise manifests as trap-depth-dependent broadening. The solid orange curve is a fit of a model in which the polarization noise is Gaussian distributed over the scan-and-fit duration (see text), yielding a standard deviation of 18(3)~kHz (3.0(4)~mrad) at the nominal trap depth.}
    \label{fig:phaseNoise}
\end{figure}

We conclude our study with an examination of polarization stability, with a focus on noise introduced by phase fluctuations of the active SLM. We carry out three complementary investigations: (i) optical polarimetry, (ii) atomic resonance stability, and (iii) atomic spectral broadening. 

We monitor fluctuations in linear polarization using a polarimeter comprising a polarizing beam splitter (PBS) and two photodiodes (PDs), as shown in Fig.~\ref{fig:phaseNoise}a. We generate all 13 tweezers and program a uniform phase mask on the SLM. The linear polarization angle is calculated from the PD voltages. Over a 24-hour period, we record PD traces at a sampling rate of $12.5~{\rm kHz}$, interleaving measurements where the tweezers are extinguished to measure the background. We record a reference measurement with the SLM off over a separate 24-hour window. We compute the periodogram from the time series of the linear polarization angle to obtain the power spectral density (PSD), which characterizes the polarization-angle noise in the frequency domain. We observe that the SLM introduces ${\approx}10$~dB of excess noise, most prominently in the Fourier frequency range from $10^{-3}$~Hz to $10^{2}$~Hz. Sharp peaks at $1.35~{\rm kHz}$, $2.71~{\rm kHz}$, and $5.41~{\rm kHz}$ correspond to harmonics of the SLM's dc balancing frequency. The PSDs for the two cases (SLM active versus inactive) coincide below $1~\rm mHz$, where the drift in both traces correlates strongly with changes in the ambient humidity and temperature at these slow timescales. Integrating the PSD across the measured range for the 24-hour datasets yields array-averaged standard deviations at the level of ${\approx}5~\rm mrad$; we attribute the small difference between the active and inactive SLM states to day-to-day variation in environmental drift.

Slow, environmentally driven polarization fluctuations highlight the need for active feedback to keep long-term drifts under control. To this end, we implement closed-loop polarization stabilization via site-resolved, in-situ spectroscopy of the ${}^1{\rm S}_0 \leftrightarrow {}^3{\rm P}_1$ transition. The feedback protocol uses the prior calibration of $\nu$ and $\theta$ versus SLM pixel voltage for each tweezer site, and particularly the slope $\partial\nu/\partial\theta$ near the magic angle (Sec.~\ref{sec:local}). We interrogate the transition periodically (every 20 minutes) to determine the frequency correction ($\Delta f$) required to null any accumulated differential light shift, and update the SLM superpixel phases for each site to steer the resonances to the free-space value. We apply feedback to the first eight tweezers while leaving the remaining five tweezers uncorrected as controls (Fig.~\ref{fig:phaseNoise}b). Over a 16-hour duration, the stabilized tweezers show both a smaller site-to-site spread in line center and a two-fold reduction in temporal fluctuation (rms deviation of $20~\rm kHz$, compared to $44~\rm kHz$ for the control sites). These values correspond to rms polarization angle deviations of $3.3~\rm mrad$ and $7.3~\rm mrad$, respectively, compatible with the optical polarimeter measurements above. The residual instability is primarily set by polarization drifts during the feedback cycle itself (i.e., the finite acquisition duration for site-resolved spectroscopy), statistical uncertainty in the fitted line centers from quantum projection noise, and, to a lesser extent, reference cavity drift.

Lastly, we investigate the extent to which polarization noise (including fluctuations introduced by the active SLM) operationally broadens the ${}^1{\rm S}_0 \leftrightarrow {}^3{\rm P}_1$ transition. Polarization-angle noise translates to effective frequency noise via the differential light shift slope $\partial\nu/\partial\theta$. Over the finite spectral acquisition (i.e., scan-and-fit) time, both second-scale and fast (${>}1~\rm kHz$) polarization noise broaden the observed line shape. Thus, this contribution to the linewidth is directly proportional to the tweezer depth. Other important broadening mechanisms include probe-laser power broadening, finite pulse time (120~$\mu \rm s$), laser linewidth, and the transition natural linewidth ($\Gamma$). In contrast, the latter contributions are trap-depth independent. To eliminate probe power broadening, we extrapolate the full width at half maximum (FWHM) of the carrier resonance to zero probe-laser power (Fig.~\ref{fig:phaseNoise}c, inset). Measuring the extrapolated FWHM at three trap depths (i.e., $1\times$, $0.75 \times$, and $0.5 \times$ the $0.5$-mK nominal trap depth) thus isolates the contribution from polarization noise (Fig.~\ref{fig:phaseNoise}c). We model this excess broadening as arising from frequency shifts sampled from a Gaussian distribution whose FWHM scales linearly with tweezer depth (Appendix~\ref{Appendix:FWHM}). From a fit of this model to the data, we find a standard deviation of $18(3)~\rm kHz$ at the nominal trap depth. This corresponds to a polarization angle standard deviation of $3.0(4)~\rm mrad$, consistent with the integrated PSD (Fig.~\ref{fig:phaseNoise}a).

\section{Conclusion and Outlook}

In summary, we have developed a scheme for independent linear polarization control in an optical tweezer array, employing an SLM as a site-resolved variable phase retarder. Through in-situ measurements on the ${}^1{\rm S}_0\leftrightarrow{}^3{\rm P}_1$ transition and iterative corrections, we uniformize the anisotropic tweezer light shifts to enable atom loading into the 813-nm array and efficient sideband cooling at the magic angle. We further propose a 2D zone architecture enabled by local polarization tunability and demonstrate the feasibility of inter-zone transport.  
With a smooth polarization transition zone, we observe high-survival transport between zones with orthogonal polarizations. In a separate measurement, we observe good coherence preservation for the clock qubit during transport across distinct polarization zones, with a slight contrast reduction largely attributable to atom heating, which could be further improved with a smooth transition zone.
Finally, we examine polarization-angle noise using optical polarimetry and site-resolved ${}^1{\rm S}_0\leftrightarrow{}^3{\rm P}_1$ spectroscopy, and implement real-time feedback based on atomic line-center tracking that suppresses slow polarization-angle drifts to the few-mrad level.

Our work sets the stage for further exploration of quantum metrology~\cite{kielinski2024ghz,direkci2026heisenberg,direkci2026extending,kaplan2025theory} and quantum simulation~\cite{zu2021emergent,fang2025probing,sun2026observation,Shaw2024C} with dynamically reconfigurable strontium arrays in 2D. With the demonstrated tweezer waist $w\approx10$ pixels and an adjusted tweezer spacing of $4w=40$ pixels, this method is readily applicable to a $48 \times 30$ array ($=1{,}440$ tweezers) with the SLM used throughout this work.
Reducing the demagnification could further reduce the corresponding superpixel width, therefore increasing the number of tweezers within the same field of view. As long as individual SLM pixels project to subwavelength dimensions in the tweezer plane, their discrete structure is effectively smoothed by the finite optical resolution. Ultimately, electric-field crosstalk between neighboring SLM pixels may limit independent tweezer control.

More generally, our results establish tweezer polarization as a programmable, per-site degree of freedom, alongside position and intensity. We expect this capability to find use where polarization shapes the trapping potential or the strength of atom--light coupling. Species with anisotropic polarizabilities stand to benefit directly. Beyond the application demonstrated in this work, prominent examples include divalent-atom arrays that leverage the ${}^3{\rm P}$ fine-structure manifold to encode quantum information~\cite{Unnikrishnan2024,Tao2026,Ammenwerth2025}, as well as magnetic lanthanides, whose open $4f$ shell confers uncommonly large vector and tensor polarizabilities~\cite{bloch2024anisotropic,becher2018anisotropic} on the states used for narrow-line laser cooling and imaging~\cite{bloch2023trapping,grun2024optical}. Studies of collective atom--light interactions in subwavelength arrays on transitions involving states of nonzero total angular momentum should gain from the same control~\cite{holman2026mid}. The rotational states of polar molecules likewise possess large tensor polarizabilities. Where rotational magic wavelengths~\cite{Ruttley2025,gregory2024second} are not readily realizable, magic polarization conditions have been engineered to extend rotational coherence times for tweezer-trapped molecules~\cite{Picard2025,Park2023,Burchesky2021}. In alkali-atom arrays, narrow electric-quadrupole transitions, such as the $6{\rm S}_{1/2}\leftrightarrow 5{\rm D}_{5/2}$ transition of Cs at 685~nm, present comparable opportunities for local addressing and magic-angle sideband cooling~\cite{blodgett2025narrow}. For quantum processors based on hyperfine qubits~\cite{Bluvstein2024,Bluvstein2026} or nuclear-spin qubits~\cite{muniz2025high}, Raman beams drive single-qubit rotations through a fictitious magnetic field (or vector light shift)~\cite{levine2022dispersive} whose strength depends on the decomposition of the local polarization into its spherical-basis components; applying our polarization control to local Raman beams would allow the Rabi frequencies to be tuned site by site. Polarization control may also enable us to compensate the fictitious magnetic field of tightly focused tweezers. Even for a linearly polarized input, a high-NA objective produces an unavoidable elliptical polarization profile near the focus. The resulting vector light shift degrades atom lifetime and qubit coherence~\cite{thompson_coherence_2013,garcia_improving_2018}. Our protocol could be extended to exploit the pixel-level programmability of the SLM, which allows one to write a spatially varying polarization profile across a single tweezer. In sum, by controlling both ellipticity and principal orientation, the two-SLM scheme of Fig.~\ref{fig:optics}c would allow anisotropic light shifts to be uniformized across large arrays for global operations or deliberately harnessed for site-resolved addressing.\\

\textbf{Acknowledgments:} We acknowledge helpful discussions with Elie Bataille, Kangheun Kim, Hudson Loughlin, and Gyohei Nomura. We acknowledge support from the Army Research Office, TINA QC (W911NF2410388 and W911NF261A009); the Army Research Office MURI program (W911NF2010136); U.S. Department of Energy (DE-SC0021951); AFOSR (FA9550-23-1-0625); the Institute for Quantum Information and Matter, an NSF Physics Frontiers Center (PHY-2317110); the Gordon and Betty Moore Foundation (GBMF11562); the Heising-Simons Foundation (20244852); Oratomic, Inc.; and the Broadcom Innovation Fund at Caltech. Support is also acknowledged from the U.S. Department of Energy, Office of Science, National Quantum Information Science Research Centers, Quantum Systems Accelerator (Award No. DE-SCL0000121). R.B.S.T. acknowledges support from the Caltech-Taiwan Fellowship. Y.L. and K.H.L. acknowledge support from AWS Quantum Postdoctoral Fellowship.\\

\textbf{Competing Interests:} M.E. is a co-founder, shareholder, and advisor of Oratomic. N.D.O. is a Visiting Faculty Researcher with Google Quantum AI. The authors are inventors on a patent application related to the protocols presented in this work.

\setcounter{section}{0}
\setcounter{equation}{0}
\captionsetup[figure]{labelfont={bf},name={FIG.},labelsep=period,justification=raggedright,font=small}
\renewcommand{\thesection}{Appendix \Alph{section}}
\renewcommand{\thesubsection}{\arabic{subsection}}
\renewcommand{\theequation}{\thesection\arabic{equation}}

\appendix
\section{Jones calculus of polarization}\label{Appendix:Jones}
In this section, we derive the Jones matrix representation of the optical setup shown in Fig.~\ref{fig:optics}~a,c. Here we use a two-dimensional vector to express an arbitrary pure polarization 
\begin{equation}
    \boldsymbol{\varepsilon} = v_1 {\bf H} + v_2 {\bf V} = \begin{pmatrix}
        v_1 \\ v_2
    \end{pmatrix}.
\end{equation}
The actions of the QWP and the SLM on the polarization state can be described by $2\times2$ unitary matrices (Jones matrices). Here, we assume that the SLM adds a phase $\varphi$ to the $V$ component, and the fast axis of the QWP is aligned with $({\bf H} + {\bf V})/\sqrt 2$. For illustration purposes, we assume that the SLM is transmissive with the beam normal to its plane, instead of reflective as in our experiment for this work. Then, the Jones matrices for the QWPs and the SLM are
\begin{equation}
    U_{\rm QWP} = \dfrac{1}{\sqrt 2} \begin{pmatrix}
        1 & -i \\ -i & 1
    \end{pmatrix},\ 
    U_{\rm SLM} = \begin{pmatrix}
        1 & 0 \\ 0 & {\rm e}^{{\rm i} \varphi}    
    \end{pmatrix}.
\end{equation}
Therefore, a QWP-SLM-QWP setup realizes a unitary transformation
\begin{equation}
    U_{\rm QWP}U_{\rm SLM}U_{\rm QWP} = -{\rm i} {\rm e}^{{\rm i}\frac{\varphi}{2}} \begin{pmatrix}
        \sin \dfrac{\varphi}{2} & \cos \dfrac{\varphi}{2} \\[6pt]
        \cos \dfrac{\varphi}{2} & -\sin \dfrac{\varphi}{2}
    \end{pmatrix},
\end{equation}
which functions as a half wave plate at angle $\frac{\pi}{4}-\frac{\varphi}{4}$. Hence, changing the phase by $\Delta \varphi$ on the SLM results in a polarization rotation of $\Delta \theta = \Delta \varphi/2$.

In Fig.~\ref{fig:optics}c we propose a setup that generates an arbitrary polarization state from a fixed input pure state. Since a polarization state is characterized by two parameters (ellipticity $\chi$ and the principal axis orientation $\theta$, for example), we need two control degrees of freedom. Suppose that we apply phases $\varphi_{1,\,2}$ to the two SLMs, respectively, and consider a special case where the input polarization is ${\bf H}$. Then, the first SLM controls ellipticity. This can be seen by defining the diagonal basis vectors
\begin{equation}
    {\bf D} = \dfrac{1}{\sqrt{2}} \begin{pmatrix}
        1 \\ 1
    \end{pmatrix},\
    {\bf A} = \dfrac{1}{\sqrt{2}} \begin{pmatrix}
        1 \\ -1
    \end{pmatrix},
\end{equation}
and noting that
\begin{equation}
\begin{split}
    &U_{\rm SLM}(\varphi_1)U_{\rm QWP} {\bf H}\\
    =\, &{\rm e}^{{\rm i}\left(-\frac{\pi}{4}+\frac{\varphi_1}{2}\right)}\left[\cos\left(\dfrac{\pi}{4}-\dfrac{\varphi_1}{2}\right)  {\bf D} + {\rm i} \sin\left(\dfrac{\pi}{4}-\dfrac{\varphi_1}{2}\right) {\bf A} \right]
\end{split}
\end{equation}
represents an elliptically polarized beam with ellipticity $\chi = \frac{\pi}{4} -\frac{\varphi_1}{2}$ with its principal axis aligned with $\bf D$ (when $|\varphi_1|\le \frac{\pi}{2}$). The following QWP-SLM-QWP setup then functions as a half wave plate that rotates the principal axis to $\theta = \frac{\pi}{4}-\frac{\varphi_2}{2}$.

\section{SLM phase calibration}\label{Appendix:SLMcalib}
\begin{figure}[t]
    \centering
    \includegraphics[width=1.0\linewidth]{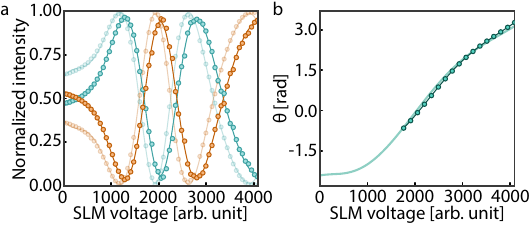}
    \caption{\textbf{SLM phase calibration.}
    \textbf{(a)} SLM phase-voltage calibration with a polarimeter. We use the polarimeter setup depicted in Fig.~\ref{fig:phaseNoise}a to measure the $s$ and $p$ components (represented by cyan and orange) after the imaging objective. Dark: all 13 tweezers; light: one tweezer. We observe a contrast of $90.6(7)\%$ and $97.8(5)\%$ respectively.
    \textbf{(b)} Polarization angle calibration with optical polarimeter and atomic spectroscopy. The light cyan curve represents the polarization we extract from the polarimeter measurement. The dark cyan curve and data points represent the angle $\theta$ between the linear polarization and the magnetic field that we extract from Fig.~\ref{fig:polarization}a.
    }
    \label{fig:calibration}
\end{figure}

We first perform an optical polarimeter measurement to determine the relation between the rotated orientation of the polarization and the applied voltage $V$ (12-bit integer sent to the controller) on an SLM pixel. We scan the voltage from $0$ to $4095$ (Fig.~\ref{fig:calibration}a) and fit the resulting angle $\theta_{\rm opt}$ to a sixth-order polynomial of the voltage (light cyan curve in Fig.~\ref{fig:calibration}b). The measurements with 13 tweezers and 1 tweezer result in a contrast of $90.6(7)\%$ and $97.8(5)\%$, respectively. The reduction of contrast when all 13 tweezers are probed is due to the polarization inhomogeneity across the array.

We determine the angle $\theta$ between the magnetic field $\bf B$ and the polarization state $\boldsymbol \varepsilon$ from spectroscopy (Fig.~\ref{fig:polarization}a). We use a trial function that maps the angle $\theta$ to a third-order polynomial of $\theta_{\rm opt}$ and fit the spectroscopy result to $V$, where the resulting calibration is shown in Fig.~\ref{fig:calibration}b (dark cyan curve and data points). Around the magic condition $V\approx 3200$, we find an angle change of $1.5~{\rm mrad}$ for every integer increment by one on $V$, which is the resolution of polarization control with a uniform superpixel phase mask. This is, in principle, device-dependent, as every manufacturer has their own controller protocols.

\section{Optimization of sideband cooling performance}\label{sec:sideband}
To optimize the sideband cooling parameters, we minimize the atomic temperature measured via trap release and recapture~\cite{tuchendler2008energy} following an 80-${\rm \mu s}$ cooling pulse from either the axial beam or one of the radial beams. We typically initialize the frequency of each beam to 100 kHz red-detuned from the free-space resonance (incidentally, this is approximately the radial sideband frequency). Additionally, we set the tweezer polarizations close to the magic angle. We observe that the axial beam yields the largest temperature reduction. Hence, the tweezer polarizations are iteratively adjusted to achieve the best cooling performance for this beam first. We then repeat this polarization optimization for the radial cooling beams, and finish with a final frequency optimization for all three cooling beams. Empirically, this procedure converges on a common operating frequency near-resonant with the radial red sideband.

\section{Polarization noise measurement via spectroscopy signal}\label{Appendix:FWHM}
In Sec.~\ref{sec:stability}, we characterize the polarization noise by measuring the differential light shift it induces on the atoms. Both the trapping intensity and the trapping light polarization affect the transition light shift. For small intensity and polarization deviations, $\delta I$ and $\delta\theta$, the change in the transition frequency is, to leading order,
\begin{equation}
    \delta \nu = \dfrac{\partial \alpha}{\partial \theta} I \delta \theta + \alpha \delta I,
\end{equation}
where $\alpha$ is the differential polarizability.
At the magic angle, $\alpha = 0$, so the second term is nulled. Therefore, we infer the polarization drift $\delta \theta$ by dividing the frequency drift $\delta \nu$ by the polarization-to-frequency conversion ratio $\partial \nu/\partial \theta = 5.95(8)~{\rm kHz/mrad}$ at the magic angle and our operating trapping intensity. This is the same ratio that we use to feedback on the SLM phase pattern based on the spectroscopy signal.

Figures~\ref{fig:phaseNoise}b and \ref{fig:phaseNoise}c provide two different measurements of the polarization noise. In Fig.~\ref{fig:phaseNoise}b, we extract the peak center position every 20 minutes. This probes the average drift of the peak center over every 20 minutes. Yet, Fig.~\ref{fig:phaseNoise}c focuses on the carrier peak width, which encodes more information. The peak width is determined by the following noise sources:
\begin{enumerate}[label=(\roman*)]
    \item the natural linewidth $\Gamma$,
    \item the probe laser linewidth and the frequency drift,
    \item the trap light shift due to an atom sampling different trapping laser intensities,
    \item the probe laser power broadening,
    \item the polarization-induced fast trap light shift fluctuation during the probe period and the slow drift among experimental shots.
\end{enumerate}
Among these effects, (iii) is nulled to the first order because we tune all tweezer polarizations to the magic condition; (i) and (ii) are independent of the probe laser power or the trap depth; (iv) only depends on the probe laser power; (v) only depends on the trap depth. Therefore, we measure the peak width dependence on the trap depth and the laser power, and fit to a model to extract the polarization-induced trap light shift fluctuation. We take data at three different trap depths under various probe laser powers. Here, we assume the laser noise has a Lorentzian line shape (with linewidth $\Gamma_{\rm laser}$) and the polarization-induced noise is a stochastic Gaussian process. We denote the standard deviation of the Gaussian noise by $\sigma$, which is proportional to the trapping laser intensity (trap depth). We perform a simulation of this stochastic process and extract the full width at half maximum (FWHM) of the carrier peak. We find the following model provides a residual bounded by $0.07\%$ under our experimental parameters:
\begin{equation}
    {\rm FWHM} = aW+\sqrt{(1-a)^2W^2 + c\sigma^2},
\end{equation}
where $a=0.57, c=3.40$, and 
\begin{equation}
    W=\sqrt{(\Gamma + \Gamma_{\rm laser})((1+s)\Gamma + \Gamma_{\rm laser})}.
\end{equation}
Note that in these equations, ${\rm FWHM}$, $W$, $\sigma$, $\Gamma$ and $\Gamma_{\rm laser}$ are all angular frequencies. By first extrapolating to $s=0$ and then fitting the widths at different trap depths, we extract $\sigma$ at nominal operating trap depth being $2\pi \times 18(3)~{\rm kHz}$. Dividing by $\partial \nu/\partial \theta$, we obtain an estimate of the polarization fluctuation of $3.0(4)~{\rm mrad}$.

\bibliography{library}

\end{document}